\documentstyle[preprint,aps]{revtex}
\begin{document}
\draft
\title{Alternative derivation of the correspondence between
Rindler and Minkowski particles}
\author{George E.A. Matsas}
\address{Instituto de F\'\i sica Te\'orica,
         Universidade Estadual Paulista\\
         Rua Pamplona 145\\
         01405-900-S\~ao Paulo, S\~ao Paulo\\
         Brazil}

\maketitle
\begin{abstract}
We develop an alternative derivation of Unruh and Wald's seminal
result that the absorption of a Rindler particle by a detector
as described by uniformly accelerated observers corresponds to
the emission of a Minkowski particle as described by inertial
observers. Actually, we present it in  an inverted
version, namely, that the emission of a Minkowski particle
corresponds in general to either the emission or the absorption of a
Rindler particle.
\end{abstract}
\pacs{04.62.+v, 04.70.Dy}

We present a brute force, but straightforward method to reobtain
Unruh and Wald's  seminal result \cite{UW} that the
{\em absorption} of a Rindler particle by a detector as described by
uniformly accelerated observers corresponds to the {\em emission} of a
Minkowski particle as described by inertial observers.
Actually, we present it in an inverted version, namely,
that the  emission of a
Minkowski particle in the inertial vacuum will correspond in
general to a
thermal  emission or absorption of a Rindler particle.
In the particular
case where the source is a uniformly accelerated Unruh--DeWitt
detector \cite{U,D}, the  emission
of a Minkowski particle will uniquely correspond to the  absorption
of a Rindler particle. We believe that our approach may be
particularly useful in understanding the behavior of realistic
sources following arbitrary worldlines.
We assume natural units  $(\hbar=c=k_B=1)$, and an
n-dimensional Minkowski spacetime with signature $(+ - -
... -)$ for sake of generality.

In order to capture only the essential features of a quantum
device without making use of some particular detector, let us
motivate the use of complex currents. The current describing
the excitation of a DeWitt--like detector \cite{D} can be
defined as
\begin{equation}
j(\tau) = \langle E \vert \hat m(\tau ) \vert E_0 \rangle ,
\label{j}
\end{equation}
where $\hat m(\tau )$ is the monopole which represents the detector,
$\tau$ is its proper time, and $\Delta E=E-E_0$ is the
energy gap of the detector.
In the Heisenberg picture the monopole is time--evolved as
$\hat m(\tau) =e^{i H_0 \tau} \hat m(0) e^{-i H_0 \tau}$,
where $ H_0$ is the free Hamiltonian of the detector. Thus,
current (\ref{j}) is clearly complex since
$j(\tau) \propto e^{i\Delta E \tau}$.  This
is the reason why we will consider here arbitrary complex currents.
These currents should be interpreted as describing the
transition of a non--necessarily pointlike quantum device
following an arbitrary worldline. For sake of simplicity, we couple
our complex current to a real massless scalar field
through the interaction operator
\begin{equation}
\hat{\cal S} =
\int d^n x \sqrt{\vert g(x) \vert } \; j(x) \hat \phi (x) .
\label{S1}
\end{equation}
Concluded the preliminaries,
let us begin by computing the emission rate of Minkowski particles
in the inertial frame. The emission amplitude of a Minkowski particle as
calculated in the inertial frame is
\begin{equation}
^M{\cal A}_{\mbox{\scriptsize em}}
(k,{\bf k}) =
              _M\langle  {\bf k} \vert \hat{\cal S}  \vert 0
\rangle_M ,
\label{MAEM}
\end{equation}
where $k= \vert {\bf k} \vert$.
Expanding the scalar field $\hat \phi(x)$
in terms of positive and negative energy modes with respect to
inertial observers \cite{BD}
\begin{equation}
\hat \phi (x^\mu ) = \int \frac{d^{n-1} {\bf k}}{ \sqrt{2 k (2 \pi)^{n-1}}}
(\hat a_{\bf k}^M e^{-ik_\mu x^\mu} + {\rm H.c.} ),
\label{FIXMU}
\end{equation}
we express from (\ref{S1}) the emission amplitude (\ref{MAEM}) as
\begin{equation}
^M{\cal A}_{\mbox{\scriptsize em}}
(k,{\bf k}) = \left[ 2 k (2 \pi)^{n-1}\right]^{-1/2} \int d^n x
j(x) e^{i k_\mu x^\mu} ,
\label{MAEM1}
\end{equation}
where $x^\mu \equiv (t,x,y^{2\leq i \leq n-1})$.
Thus, the emission probability of a Minkowski particle as described by
inertial observers is
\begin{equation}
^M{\cal P}_{\mbox{\scriptsize em}}
= \int d^{n-1} {\bf k} \,
\vert ^M{\cal A} {\mbox{\scriptsize em}(k, {\bf k} )} \vert^2 ,
\label{PM}
\end{equation}
where $^M{\cal A}_{em}$ is given in (\ref{MAEM1}).

Next, we aim to express the  emission rate of Minkowski
particles (\ref{PM}) in terms of the  emission and  absorption
 of Rindler  particles. For this purpose, it is convenient to introduce
Rindler coordinates $(\tau, \xi, y^{2 \leq i\leq n-1})$.
These coordinates are related with Minkowski coordinates
$(t,x,y^{2\leq i\leq n-1})$  by
\begin{equation}
t = \frac{e^{a\xi}}{a}\sinh a\tau, \;
x = \frac{e^{a\xi}}{a}\cosh a\tau.
\label{RC}
\end{equation}
A worldline defined by $\xi, y^{2\leq i\leq n-1} =$ const describes an
observer with constant proper acceleration $a
e^{-a\xi}$. We will denominate these observers {\em Rindler} observers.
The natural manifold to describe Rindler observers
is the Rindler wedge, {\em i.e.} the portion of the Minkowski space defined
by $x > \vert t \vert$. It is crucial for our purposes that the current
$j(x^\mu)$ be confined inside this wedge.
The Rindler wedge is a globally hyperbolic spacetime
in its own right, with a Killing horizon at $x=\pm t$ ($\tau =
\pm \infty $) associated with the boost Killing field $\partial_
\tau$ at its boundary.
Using (\ref{RC}),
the Minkowski line element restricted to the Rindler wedge is
\begin{equation}
ds^2 = e^{2a\xi} (d\tau^2 - d\xi^2) - \sum_{i=2}^{n-1} (d {y^i})^2 .
\label{DS}
\end{equation}
Solving the massless Klein-Gordon equation $\Box \phi =0$ in the
Rindler wedge, we obtain a complete set of Klein-Gordon
orthonormalized functions \cite{F}
\begin{equation}
u_{\omega {\bf k}_\bot} (x^\mu) =\left[ \frac{2 \sinh \pi \omega/a}{
(2\pi)^{n-1} \pi a}\right]^{ \frac{1}{2}} K_{i\omega/a}
\left(\frac{k_\bot}{a} e^{a\xi } \right)
e^{i {\bf k}_\bot {\bf y} - i \omega \tau} ,
\label{U}
\end{equation}
where $k_\bot \equiv \vert {\bf k}_\bot \vert =
\sqrt{\sum_{i=2}^{n-1} (k_{y^i})^2}$, $\omega$  is the frequency
of the Rindler mode, and $K_{i\lambda }(x)$ is the MacDonald function.
Using (\ref{U}), we express the scalar field in the uniformly
accelerated frame in terms of
positive and negative frequency modes with respect to the boost
Killing field $\partial_\tau$
\begin{equation}
\hat \phi (x^\mu ) = \int d^{n-2} {\bf k}_\bot
\int_{0}^{+\infty} d\omega
\left\{{\hat a^R}_{ \omega {\bf k}_\bot}
       u_{\omega {\bf k}_\bot }
       (x^\mu ) + {\rm H.c.} \right\},
\label{FIXMU2}
\end{equation}
where $\hat a^R_{\omega {\bf k}_\bot}$ is the annihilation operator of
Rindler particles, and obeys the usual commutation relation
\begin{equation}
\left[\hat a^R_{ \omega {\bf k}_\bot } ,
\hat a^{R \dagger}_{ \omega' {\bf k}'_\bot } \right ] =  \delta (\omega -
\omega')
\delta  ({\bf k}_\bot - {\bf k}'_\bot) .
\label{CR}
\end{equation}

It is possible now to Fourier analyze the current $j(x^\mu )$ in
terms of Rindler modes. We define its {\em Rindler--Fourier}
transform as
\begin{equation}
\tilde \jmath_R ( \omega_0, \omega, k_\bot) \equiv
\int d^n x \sqrt{\vert g(x)\vert } j(x^\mu ) u^*_{\omega {\bf k}_\bot}
e^{-i(\omega - \omega_0) \tau} .
\label{RFT}
\end{equation}
This relation can be easily inverted
\begin{equation}
j(x^\mu) = \frac{e^{-2a\xi}}{\pi}\int d^{n-2} {\bf k}_\bot
        \int_0^{+\infty} d\omega \omega
        \int_{-\infty}^{+\infty} d\omega_0
\tilde{\jmath}_{R}(\omega_0, \omega , k_\bot )
u_{\omega {\bf k}_\bot }(x^\mu ) e^{i(\omega-\omega_0)\tau}
\label{JX2}
\end{equation}
by using the completeness relation \cite{HM}
\begin{equation}
 \int_0^{+\infty} d\omega \omega \sinh \frac{\pi \omega}{a} \;
 K_{i\omega/a}\left(\frac{k_{\bot}}{a}e^{a\xi}\right)
K_{i\omega/a}\left(\frac{k_{\bot}}{a}e^{a\xi'}\right)
=\frac{\pi^2 a}{2}\delta(\xi-\xi') .
\label{CR2}
\end{equation}

In order to relate the particle emission and absorption rates
in both reference frames,
we substitute (\ref{JX2}) and (\ref{U}) in (\ref{MAEM1}). The
$ y^i$ and ${\bf k}_\bot$ integrals are easy to perform, while
the $\tau$ and $\xi$ integrals can be solved by  noting that
(use the change of variables $\eta =
e^{a\tau}$ in conjunction with Eq. 3.471.10  of Ref. \cite{GR})
\begin{equation}
\int_{-\infty}^{+\infty} d\tau e^{i(kt-k_x x - \omega_0 \tau)} =
\frac{2}{a} \left[\frac{k+k_x}{k-k_x}\right]^{-i\omega_0/2a}
e^{\pi \omega_0/2a} K_{i\omega_0/a}
\left( \frac{k_\bot}{a} e^{a\xi} \right) ,
\label{I1}
\end{equation}
and by using the following orthonormality relation \cite{G,HMS}:
\begin{equation}
 \int_{-\infty}^{+\infty} d\xi\; K_{i\omega/a}\left(\frac{k_{\bot}}{a}e^{a\xi}
\right)
K_{i\omega'/a}\left(\frac{k_{\bot}}{a}e^{a\xi}\right)
=\frac{\pi^2 a}{2\omega\sinh(\pi\omega/a)}
\delta(\omega-\omega')
\label{I2}
\end{equation}
for $ \omega, \omega' \in {\bf R_+}$. With this procedure we
reduce the emission amplitude (\ref{MAEM}) to
\begin{eqnarray}
^M{\cal A}_{\mbox{\scriptsize em}} (k,{\bf k}) =
\frac{1}{(4 \pi a k)^{1/2}}
\int_0^{+\infty }
& &
\frac{d\omega }{\sinh^{1/2} (\pi \omega/a)}\;
\left\{
\tilde \jmath_{R} (\omega , \omega , {\bf k}_\bot )
\left(\frac{k + k_x}{k - k_x} \right)^{- i\omega /2a}
e^{\pi \omega/2a}
\right.
\nonumber \\
& &
\left.
+ \tilde \jmath_{R} (-\omega , \omega , {\bf k}_\bot )
\left(\frac{k + k_x}{k - k_x}  \right)^{+i\omega /2a}
e^{- \pi \omega/2a}
\right\} .
\label{MAEM3}
\end{eqnarray}

Our next task will be to interpret $\tilde \jmath_R (\pm
\omega, \omega, {\bf k} _\bot)$ in terms of the emission
and absorption amplitudes of Rindler particles
\begin{equation}
^R{\cal A}_{\mbox{\scriptsize em}}
= _R\langle  {\omega {\bf k}_\bot} \vert \; \hat{\cal S} \vert 0
\rangle_R ,\;\; ^R{\cal A}_{\mbox{\scriptsize abs}}
= _R\langle 0 \vert \hat{\cal S} \vert \omega  {\bf k}_\bot \rangle_R
\label{acima}
\end{equation}
respectively, where $\hat{\cal S}$ is given in (\ref{S1}).
Using explicitly (\ref{S1}) and (\ref{FIXMU2}) in (\ref{acima})
we obtain
\begin{equation}
\tilde \jmath_R (\omega, \omega, {\bf k}_\bot) =
^R{\cal A}_{\mbox{\scriptsize em}} (\omega, {\bf k}_\bot) ,\;\;\;
\tilde \jmath_R (-\omega, \omega, {\bf k}_\bot) =
^R{\cal A}_{\mbox{\scriptsize abs}} (\omega, - {\bf k}_\bot) .
\end{equation}
As a consequence, we can express the Minkowski particle emission
amplitude (\ref{MAEM3}) as
\begin{eqnarray}
^M{\cal A}_{\mbox{\scriptsize em}} (k,{\bf k}) =
\frac{1}{(4 \pi a k)^{1/2}}
\int_0^{+\infty }
& & \frac{d\omega }{\sinh^{1/2} (\pi \omega/a)}\;
\left\{
^R{\cal A}_{\mbox{\scriptsize em}} (\omega, {\bf k}_\bot)
\left(\frac{k + k_x}{k - k_x} \right)^{- i\omega /2a}
e^{\pi \omega/2a}
\right.
\nonumber \\
& &
\left.
+ ^R{\cal A}_{\mbox{\scriptsize abs}} (\omega, -{\bf k}_\bot)
\left(\frac{k + k_x}{k - k_x}  \right)^{+i\omega /2a}
e^{- \pi \omega/2a}
\right\} .
\label{MAEM4}
\end{eqnarray}
Now, it is useful to introduce the following representation of
the delta function
\begin{equation}
\frac{1}{2\pi a} \int_{-\infty}^{+\infty} \frac{dk_x}{k}
\left[\frac{k+k_x}{k-k_x}\right]^{-i(\omega -\omega')/2a}
= \delta (\omega -\omega') ,
\label{DR}
\end{equation}
where $k=\sqrt{k_x^2 + k_\bot^2}$.
(This delta function representation can be cast in the more familiar
form
$ \int_{-\infty}^{+\infty}
dK e^{-iK (\omega -\omega')}
= 2\pi \delta (\omega -\omega')
$
after the change of variables $ k_x \to K =
\ln [(k+k_x)/(k-k_x)]$.)
Finally, introducing
(\ref{MAEM4}) in (\ref{PM}), and using (\ref{DR}) to perform
the integral in $k_x$ we are able to express the total emission rate of
Minkowski particles in its final form as
\begin{equation}
^M{\cal P}_{\mbox{\scriptsize em}} =
^R{\cal P}_{\mbox{\scriptsize em}} +
^R{\cal P}_{\mbox{\scriptsize abs}} ,
\label{PEM6}
\end{equation}
where
\begin{equation}
^R{\cal P}_{\mbox{\scriptsize em}} =
\int d^{n-2} {\bf k}_\bot  \int_{0}^{+\infty} d\omega
\vert
^R{\cal A}_{\mbox{\scriptsize em}} (\omega, {\bf k}_\bot)
\vert^2
\left( 1 + \frac {1}{e^{2\pi \omega /a} - 1} \right),
\label{FIM2}
\end{equation}
and
\begin{equation}
^R{\cal P}_{\mbox{\scriptsize abs}} =
\int d^{n-2} {\bf k}_\bot  \int_{0}^{+\infty} d\omega
\vert ^R{\cal A}_{\mbox{\scriptsize abs}} (\omega, - {\bf k}_\bot)
\vert^2
\frac {1}{e^{2\pi \omega /a} - 1} .
\label{FIM3}
\end{equation}
The thermal factors which appear in (\ref{FIM2}) and (\ref{FIM3})
are in agreement with the fact that the Minkowski vacuum
corresponds to a thermal state with respect to uniformly
accelerated observers \cite{U}. The physical content of
(\ref{PEM6}) combined with (\ref{FIM2}) and (\ref{FIM3})
can be summarized as follows: {\em The emission of a
Minkowski particle in the vacuum with some fixed transverse
momentum ${\bf k}_\bot$
as described by an inertial observer {\em (\ref{PM})} will correspond
either to the emission of a Rindler particle with
the same transverse momentum ${\bf k}_\bot$, or to the
absorption of a Rindler particle with transverse momentum
$-{\bf k}_\bot$ from the Davies-Unruh thermal bath
as described by uniformly accelerated observers.} Notice that the
conservation of transverse momentum in both frames appears
naturally enclosed in this result. This is mandatory since the
transverse momentum is invariant under boosts.
In the particular case where the
source is a uniformly accelerated Unruh-DeWitt detector,
$^R{\cal A}_{\mbox{\scriptsize em}}=0$ implying
that in this case the absorption of a Rindler particle corresponds
uniquely to the emission of a Minkowski particle. However, as
seen above, in
more general situations where the detector is switched on/off
\cite{HMP} or follows some arbitrary worldline, the excitation of
the detector usually associated with the {\em absorption} of a Rindler
particle can be also associated with the {\em emission} of a
Rindler particle. This is also in agreement with energy conservation
arguments.

\begin{flushleft}
{\bf{\large Acknowledgements}}
\end{flushleft}

I am really indebted to Atsushi Higuchi for various
enlightening discussions. I am also very grateful to
Ulrich Gerlach for critically reading a previous version of this
manuscript.
This work was partially supported by Conselho Nacional de
Desenvolvimento Cient\'\i fico e Tecnol\'ogico.

\end{document}